\documentclass[conference]{IEEEtran}
\IEEEoverridecommandlockouts
\usepackage[T1]{fontenc}
\usepackage{cite}
\usepackage{amsmath,amssymb,amsfonts}
\usepackage{mathtools}
\usepackage[dvipdfmx]{graphicx}
\usepackage{textcomp}
\usepackage{xcolor}
\usepackage{algorithm}
\usepackage{algorithmic}
\usepackage{latexsym}
\usepackage{ascmac}
\usepackage[caption=false,font=footnotesize]{subfig} 
\def\BibTeX{{\rm B\kern-.05em{\sc i\kern-.025em b}\kern-.08em
    T\kern-.1667em\lower.7ex\hbox{E}\kern-.125emX}}

\newcommand{\ut}{{\cal UT}}
\newcommand{\luttp}{{\cal LU\!T\!T\!P}}

\newcommand{\mybk}{\hspace{-0.11em}}
\newcommand{\cs}{C{\mybk}S}
\newcommand{\CSs}{C{\mybk}S{\mybk}s}
\newcommand{\vs}{V{\mybk}S}
\newcommand{\ins}{I{\mybk}n{\mybk}S}
\newcommand{\h}{{\cal H}}

\newtheorem{definition}{Definition}

\newtheorem{theorem}{Theorem}
\newtheorem{example}{Example}

\begin{document}

\title{
Efficient Pattern Matching in Unordered Term Tree Patterns with Height Constraints
\thanks{This is an author preprint of a paper presented at the 20th International Conference on E-Service and Knowledge Management (ESKM 2025), IIAI-AAI 2025. The paper was accepted for inclusion in the conference proceedings published by IEEE Computer Society Conference Publishing Services (CPS).}
\thanks{This work was partially supported by JSPS KAKENHI Grant Numbers JP21K12021, JP24K15074, JP24K15090.}
}

\author{
    \IEEEauthorblockN{
        Shintaro Matsushita\IEEEauthorrefmark{1}\IEEEauthorrefmark{3},
        Takayoshi Shoudai\IEEEauthorrefmark{1}\IEEEauthorrefmark{4}, and
        Yusuke Suzuki\IEEEauthorrefmark{2}
    }
    \IEEEauthorblockA{
        \IEEEauthorrefmark{1}Department of Computer Science and Engineering, Fukuoka Institute of Technology, Fukuoka 811-0295, Japan\\
        Email: \IEEEauthorrefmark{3}mfm25113@bene.fit.ac.jp, \IEEEauthorrefmark{4}shodai@fit.ac.jp
    }
    \IEEEauthorblockA{
        \IEEEauthorrefmark{2}Faculty of Information Sciences, Hiroshima City University, Hiroshima 731-3194, Japan\\
        Email: y-suzuki@hiroshima-cu.ac.jp
    }
}

\maketitle

\begin{abstract}
  Unordered trees appear in applications where the order among child vertices is insignificant, such as abstract syntax trees and chemical structures. To describe patterns in such trees, we propose unordered term tree patterns, which employ height-constrained variables that restrict trunk length and subtree height. We formalize the pattern matching problem between an unordered term tree pattern and an unordered tree, and present an $O(N \cdot \max\{nD^{3/2}, \mathcal{S}\})$-time algorithm, where $n$ and $N$ are the numbers of vertices in the pattern and tree, $D$ is the maximum vertex degree, and $\mathcal{S}$ is the sum of trunk constraints. Computational results show that the algorithm runs efficiently in practice.
\end{abstract}

\begin{IEEEkeywords}
  height constrained tree pattern, polynomial-time algorithm, pattern matching algorithm, membership problem
\end{IEEEkeywords}


\section{Introduction}\label{sec:intro}

Unordered trees, where the order of sibling vertices is not significant, appear in various domains such as glycan structures, chemical compounds, and abstract syntax trees. To analyze such data, many studies have explored pattern discovery and matching techniques for unordered trees.
Earlier research focused on mining frequent subtrees~\cite{zaki2005,chehreghani2011} and addressing structural containment and inclusion problems~\cite{gottlob05,bille05}. However, these approaches are often limited in capturing flexible and complex structures that may arise in real-world data.
To overcome these limitations, several methods have been proposed to improve both the expressiveness and efficiency of pattern matching and mining~\cite{zhang2015new,akutsu2021,wu2021}. In addition, surveys such as~\cite{livi2013graph} have emphasized the need for more general frameworks for analyzing tree patterns.

As one such framework, Shoudai et al.~\cite{shoudai-ieice2018} introduced \emph{unordered term tree patterns}, in which variables can represent entire subtrees. This model was later extended with \emph{height-constrained variables} (HC-variables)~\cite{shoudai-ieice2017}, which allow control over both the trunk length and the height of substituted subtrees.
In this paper, we address the pattern matching problem for \emph{linear unordered term tree patterns with HC-variables}, where linearity means that all variables have distinct labels. We propose a polynomial-time algorithm that solves the membership problem of determining whether an unordered tree $T$ can be obtained from a given pattern $t$ by replacing each HC-variable with an appropriate subtree.

\begin{figure}[t]
  \begin{center}
    \includegraphics[scale=.33]{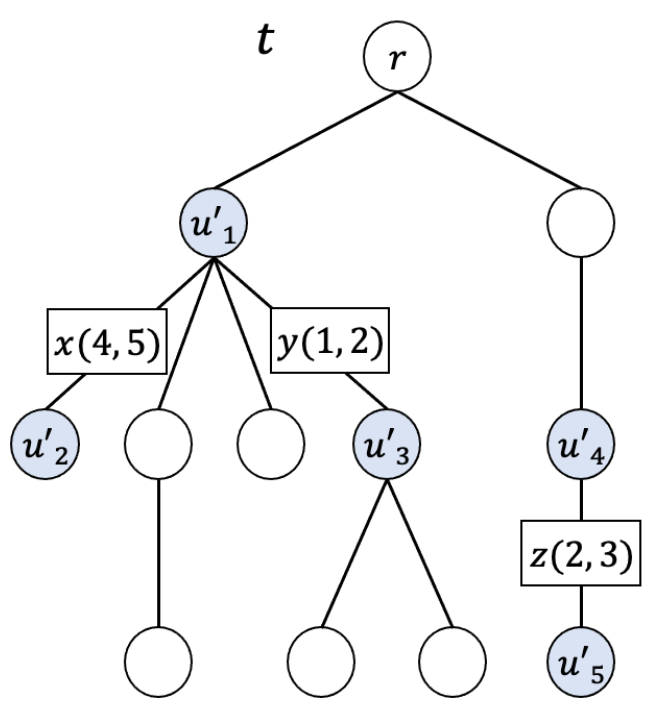}
  \end{center}
  \caption{
    An unordered term tree pattern $t$ with HC-variables.
    Circles represent vertices, and squares denote variables with parent and child ports.
    A square labeled $x(i,j)$ indicates an $(i,j)$-HC-variable.
  }
  \label{fig:ex-u-term-tree1}
\end{figure}

\begin{figure}[t]
  \begin{center}
    \includegraphics[scale=.33]{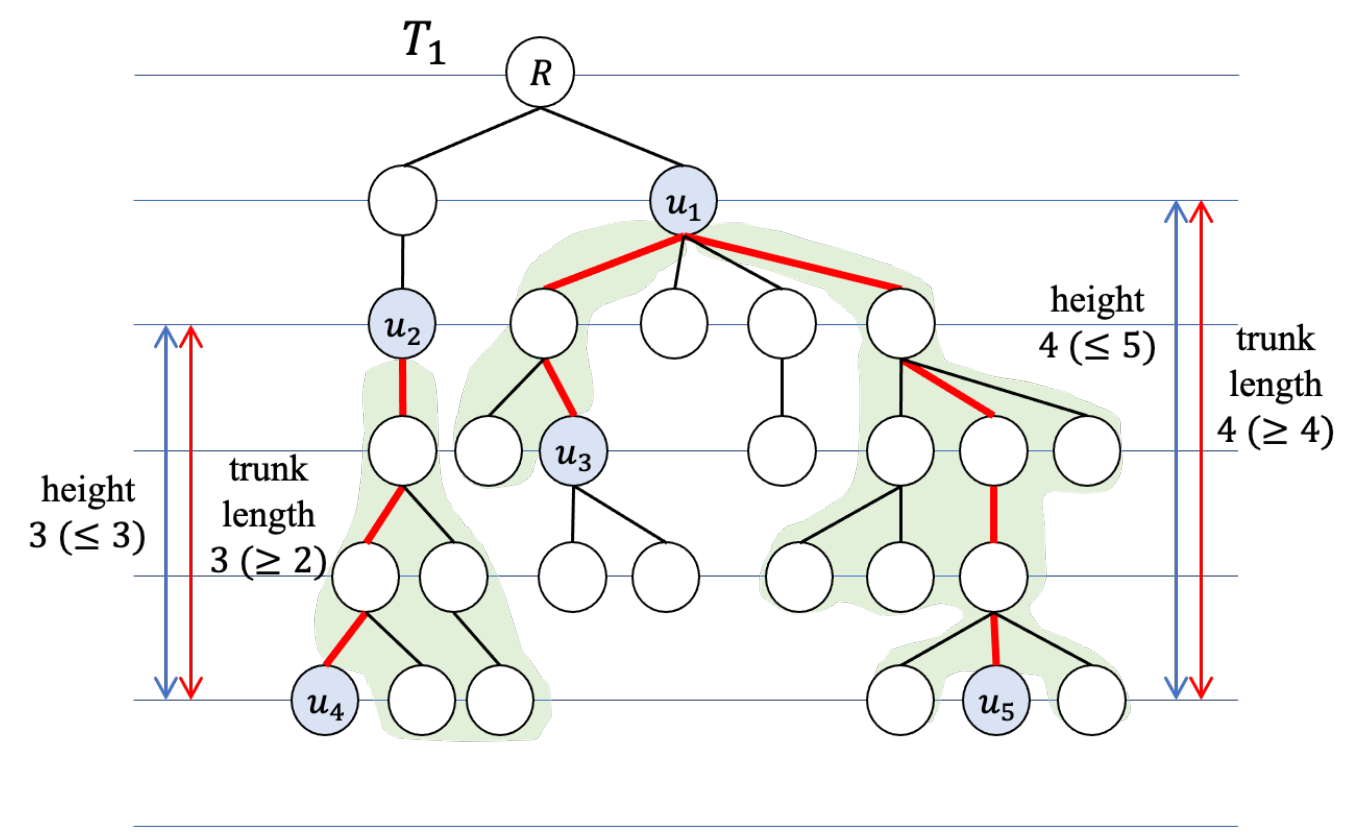}\\[1ex]
    \includegraphics[scale=.33]{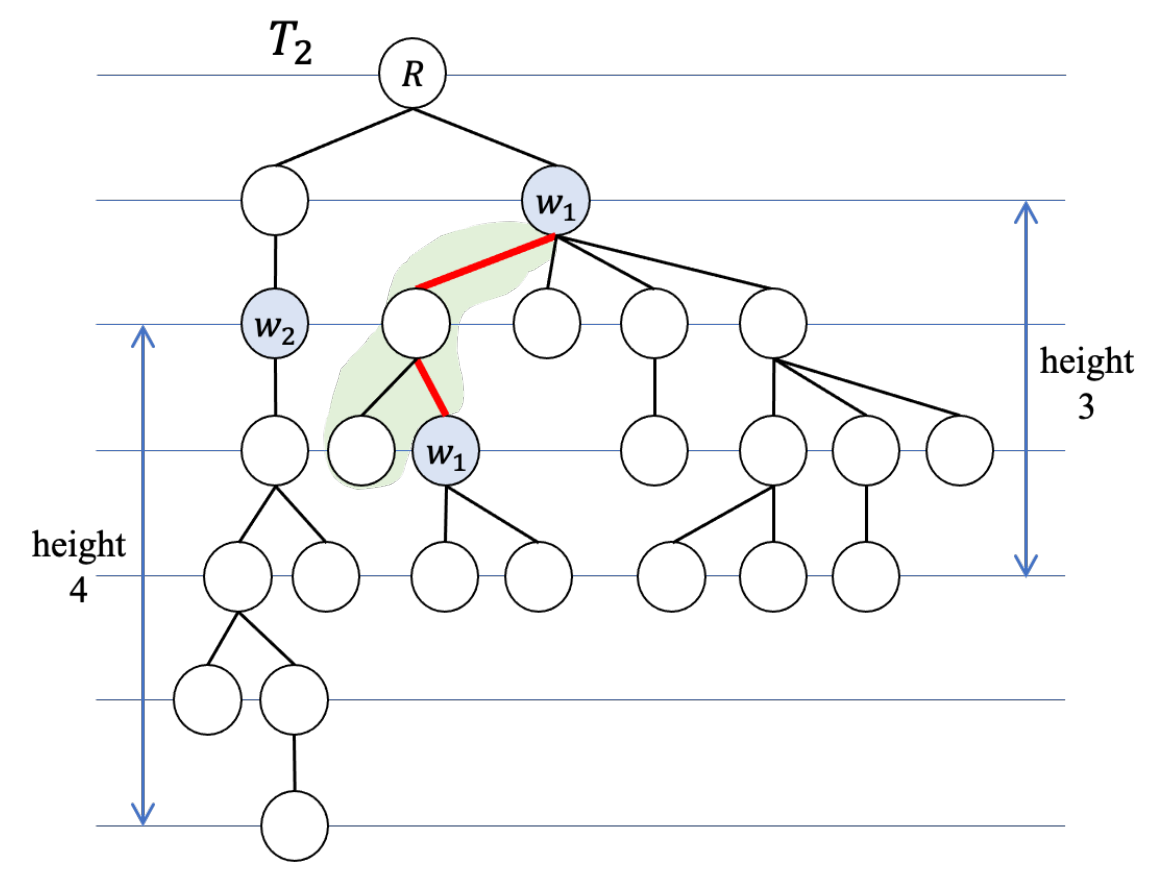}
  \end{center}
  \caption{
    Unordered trees $T_1$ and $T_2$:
    The pattern $t$ in Fig.~\ref{fig:ex-u-term-tree1} matches $T_1$ but not $T_2$.
  }
  \label{fig:ex-u-trees}
\end{figure}

An example of such a pattern $t$ is shown in Fig.~\ref{fig:ex-u-term-tree1}, which includes three HC-variables: $x(4,5)$, $y(1,2)$, and $z(2,3)$. Fig.~\ref{fig:ex-u-trees} shows two example unordered trees, $T_1$ and $T_2$. The tree $T_1$ can be derived from $t$ by replacing each HC-variable with a subtree satisfying the corresponding constraints. In contrast, $T_2$ does not satisfy the constraints and therefore does not match the pattern.

The proposed algorithm runs in $O(N \cdot \max\{n D^{3/2}, \mathcal{S}\})$ time, where $n$ and $N$ denote the number of vertices in $t$ and $T$, respectively, $D$ is the maximum degree of internal vertices in $T$, and $\mathcal{S}$ is the total trunk constraint value of the HC-variables in $t$.
This work extends previous studies on unordered tree pattern matching by incorporating structural flexibility and global height constraints using HC-variables. We believe that the proposed method will be useful for data mining applications that require expressive yet computationally efficient pattern matching in unordered trees.

\section{Preliminaries}\label{sec:preliminary}

We define unordered term tree patterns of dimension $2$. The general definition for any dimension can be found in \cite{shoudai-ieice2018}.

For a set $S$, the number of elements in $S$ is denoted by $|S|$.
Let $\Sigma$ and $\Lambda$ be finite alphabets.
$\ut_{\Sigma,\Lambda}$ denotes the set of all unordered trees in which the vertices are labeled with symbols from $\Sigma$ and the edges are labeled with symbols from $\Lambda$.
Symbols in $\Sigma$ are called {\it vertex labels\/}, and those in $\Lambda$ are called {\it edge labels\/}.
Additionally, let $X$ be an infinite alphabet, whose symbols are called {\it variable labels\/}.
We assume that $(\Sigma\cup\Lambda)\cap X=\emptyset$.

\smallskip

\begin{definition}[Linear unordered term tree pattern]
Let $T=(V_{T},E_{T})$ be an unordered tree.
Let $E_{t}$ and $H_{t}$ be a partition of $E_{T}$, i.e., $E_{t}\cup H_{t}=E_{T}$ and $E_{t}\cap H_{t}=\emptyset$.
An {\it unordered term tree pattern\/} is a triple $t=(V_{t},E_{t},H_{t})$ obtained from $T$ by using a partition of $E_{T}$, where $V_{t}$ is equal to $V_{T}$.
Elements in $V_{t}$, $E_{t}$ and $H_{t}$ are called a {\it vertex\/}, an {\it edge\/} and a {\it variable\/}, respectively.
All vertices, edges, and variables are labeled with labels from alphabets $\Sigma$, $\Lambda$, and $X$, respectively. 
The root of $t$ is the root of $T$ and the leaves of $t$ are the leaves of $T$.
An unordered term tree pattern $t=(V_{t},E_{t},H_{t})$ is called {\it linear\/} if all variables in $H_{t}$ have mutually distinct variable labels in $X$.
In this paper, we will focus on exclusively on linear unordered term tree patterns.
Hence, it is assumed that all unordered term tree patterns discussed are linear.
\end{definition}

\smallskip

\indent
Here we use the same terminologies as in graph theory.
For an unordered term tree pattern $t=(V_{t}, E_{t}, H_{t})$ and its vertices $v_1$ and $v_i$, a {\it path\/} from $v_1$ to $v_i$ is a sequence $v_1,v_2,\ldots,v_i$ of distinct vertices of $t$ such that for any $j$ with $1\leq j < i$, there exists an edge or a variable that consists of $v_j$ and $v_{j+1}$.
The length of a path is defined as the sum of the number of edges and the number of variables in it.
If there is an edge or a variable that consists of $v$ and $v'$ such that $v$ lies on the path from the root to $v'$, then $v$ is said to be the {\it parent\/} of $v'$ and $v'$ is a {\it child\/} of $v$. 
For an unordered term tree pattern $t$, whenever a vertex $v \in V_{t}$ is the parent of a vertex $v' \in V_{t}$, the edge connecting $v$ to $v'$ is represented by the ordered pair $(v, v')$, and the variable connected $v$ to $v'$ is represented by the list $[v,v']$.
Specifically, we call $v$ the {\it parent port\/} of $[v,v']$ and $v'$ the {\it child port\/} of $[v,v']$.
Hereafter, we simply refer to a vertex that is the child port of a variable as a {\it child port\/}.
The {\it height\/} of an unordered term tree pattern $t$ is defined as the maximum length of any path from the root to a leaf.
A vertex that is not a leaf is referred to as an internal vertex.
\smallskip

\begin{definition}[Height-constrained (HC) variables]
Let $X^\h$ be an infinite subset of $X$.
For two integers $1\leq i\leq j$, define $X^{\h(i,j)}$ as an infinite subset of $X^\h$.
It is assumed that $X^{\h}$ is the union of these subsets, i.e., $X^\h=\bigcup_{1\leq i\leq j}X^{\h(i,j)}$, and that $X^{\h(i,j)}\cap X^{\h(i',j')}=\emptyset$ for all distinct pairs $(i,j)\not=(i',j')$.
A variable label in $X^{\h(i,j)}$ is called an {\it $(i,j)$-height-constrained variable label\/} for any integers $1\leq i\leq j$.
Furthermore, a variable $[v,v']$ of an unordered term tree pattern is said to be an {\it $(i,j)$-height-constrained variable\/} (abbreviated as $(i,j)$-HC-variable) if it is labeled with an $(i,j)$-height-constrained variable label.
\end{definition}

\smallskip

\indent
$\luttp_{\Sigma,\Lambda, X^{\h}}$ denotes the set of all linear unordered term tree patterns in which vertices are labeled with vertex labels in $\Sigma$, the edges are labeled with edge labels in $\Lambda$, and the variables are labeled with height-constrained (HC) variable labels in $X^{\h}$.

\begin{definition}[Bindings for HC-variables]
Let $f$ and $g$ be unordered term tree patterns, each containing at least two vertices in $\luttp_{\Sigma,\Lambda, X^{\h}}$, and let $x$ be a variable label in $X^{\h(i,j)}$ for some integers $1\leq i\leq j$.
Consider $\sigma=[u,u']$ to be a list of two vertices in $g$, where $u$ is the root of $g$ and $u'$ is a leaf of $g$.
The length of the path between $u$ and $u'$ is referred as the {\it trunk length\/} of $\sigma$.
The notation $x\coloneqq[g,\sigma]$ is called a {\it binding\/} for $x$ if the following conditions are satisfied:
\begin{enumerate}
\item[(i)] the trunk length of $\sigma$ is at least $i$, and 
\item[(ii)] the height of $g$ is at most $j$.
\end{enumerate}
Furthermore, the trunk length of a binding $x\coloneqq[g,\sigma]$ is defined as the trunk length of $\sigma$.
\end{definition}

\smallskip

\begin{definition}[Substitutions of unordered term tree patterns]
Let $f=(V_{f},E_{f},H_{f})$ be an unordered term tree pattern in $\luttp_{\Sigma,\Lambda, X^{\h}}$.
Let $x$ be a variable label in $X^{\h(i,j)}$ for some integers $1\leq i\leq j$ and $x\coloneqq[g,\sigma]$ a binding for an unordered term tree pattern $g\in\luttp_{\Sigma,\Lambda, X^{\h}}$ and a list $\sigma=[u,u']$ of two vertices $u,u'$ in $g$.
A new unordered term tree pattern $f\{x\coloneqq[g,\sigma]\}$ is obtained by applying the binding $x\coloneqq[g,\sigma]$ to $f$ in the following way:
Let $h=[v,v']$ be an $(i,j)$-HC-variable in $f$ whose variable label is $x$.
Let $g'$ be one copy of $g$ and $w,w'$ the vertices of $g'$ corresponding to $u,u'$ of $g$, respectively.
For the variable $h=[v,v']$, we attach $g'$ to $f$ by removing $h$ from $H_{f}$ and by identifying the vertices $v,v'$ with the vertices $w,w'$ of $g'$, respectively.
The vertex labels of new vertices corresponding to $v$ and $v'$ are retained as the original labels of $v$ and $v'$ in $f$, respectively.
A {\it substitution\/} $\theta$ is a finite collection of bindings $\{x_{1}\coloneqq[g_{1},\sigma_{1}],\cdots,x_{n}\coloneqq[g_{n},\sigma_{n}]\}$, where $x_{i}$ are mutually distinct variable labels in $X^{\h}$, and each $g_i$ ($1\leq i\leq n$) has no variable label in $\{x_{1},\ldots,x_{n}\}$.
The unordered term tree pattern $f\theta$, called the {\it instance\/} of $f$ by $\theta$, is obtained by applying the all bindings $x_{i}\coloneqq[g_{i},\sigma_{i}]$ on $f$ simultaneously.
The root of the resulting unordered term tree pattern $f\theta$ is the root of $f$.
\end{definition}

We say that two unordered term tree patterns $f=(V_{f},E_{f},H_{f})$ and $g = (V_{g},E_{g},H_{g})$ are {\it isomorphic}, denoted by $f \cong g$, if there is a bijection $\varphi$ from $V_{f}$ to $V_{g}$ such that
\begin{enumerate}
\item[(i)] the root of $f$ is mapped to the root of $g$ by $\varphi$,
\item[(ii)] $\{u,v\} \in E_{f}$ if and only if $\{\varphi(u),\varphi(v)\} \in E_{g}$, and
\item[(iii)] $[u,v] \in H_{f}$ is an $(i,j)$-HC-variable if and only if $[\varphi(u),\varphi(v)] \in H_{g}$ is an $(i,j)$-HC-variable for some integers $1\leq i\leq j$.
\end{enumerate}

\begin{definition}[Unordered term tree pattern languages]
For an unordered term tree pattern $t\in \luttp_{\Sigma,\Lambda, X^{\h}}$, the {\it unordered term tree pattern language} $L(t)$ is defined as follows:
\begin{align*}
L(t) & = \{ T \in \ut_{\Sigma,\Lambda} \mid T \cong t \theta \mbox{ for a substitution }\theta \}.
\end{align*}
\end{definition}

The membership problem for $\luttp_{\Sigma,\Lambda, X^{\h}}$ is defined as follows:
In this paper, we give a polynomial time algorithm for solving the following problem for any alphabets $\Sigma$ and $\Lambda$. 

\smallskip

{\bf Membership Problem for $\luttp_{\Sigma,\Lambda, X^{\h}}$}.\par
{\bf Instance}: $t\in\luttp_{\Sigma,\Lambda, X^{\h}}$ and $T\in\ut_{\Sigma,\Lambda}$.\par
{\bf Question}: Is $T$ in $L(t)$?

\section{A Polynomial-Time Matching Algorithm for $\luttp_{\Sigma,\Lambda, X^{\h}}$}\label{sec:member-uottp}

\subsection{Algorithm {\sc Matching}}

\algsetup{linenosize=\small}
\begin{algorithm}[t]
  \floatname{algorithm}{Algorithm}
  \caption{\sc Match-$\luttp_{\Sigma,\Lambda,X^{\h}}$}\label{proc:Matching}
  {\fontsize{9}{10.5}\selectfont 
  \begin{algorithmic}[1]
   \REQUIRE an unordered term tree pattern $t=(V_t,E_t,H_t)\in\luttp_{\Sigma,\Lambda,X^{\h}}$ and an unordered tree $T=(V_T,E_T)\in\ut_{\Sigma,\Lambda}$;
  \ENSURE ``yes'' if $t$ matches $T$, otherwise ``no'';
  \smallskip
  \STATE Let $r$ and $R$ be the roots of $t$ and $T$, respectively;
  \STATE Let $Rule(t)$ be the set of matching rules of all internal vertices of $t$;
  \STATE $V\coloneqq V_{T}$;
  \FOR{each vertex $u\in V_{T}$}\label{match:beginbase}
    \IF{$u$ is a leaf of $T$}
      \STATE $\cs(u)\coloneqq\emptyset$;
      \FOR{each leaf $u'$ of $t$}
        \IF{$u'$ is a child port}
          \STATE $\cs(u)\coloneqq \cs(u)\cup \{(u',0,0)\}$;
        \ELSE
          \STATE $\cs(u)\coloneqq \cs(u)\cup \{u'\}$;
        \ENDIF
      \ENDFOR
      \STATE $\text{height}_T(u)\coloneqq 0$;
      \STATE $V\coloneqq V\setminus\{u\}$;
    \ELSE
      \STATE $\cs(u)\coloneqq\emptyset$;
      \STATE $\text{height}_T(u)\coloneqq\infty$;
    \ENDIF
  \ENDFOR\label{match:endbase}
  \WHILE{$V \not= \emptyset$}\label{match:begininductive}
    \FOR{each vertex $u\in V$}
      \STATE Let $c_{1},\ldots,c_{m}$ be all of the children of $u$;
      \IF{all $c_{\ell}$ $(1\leq \ell \leq m)$ are in $V_{T}\setminus V$}
        \STATE $\CSs\coloneqq(\cs(c_{1}),\ldots,\cs(c_{m}))$;
        \STATE $\ins(u)\coloneqq\mbox{\sc I\_Set\_Making}(u, \CSs, H_{t})$;
        \STATE $\vs(u)\coloneqq\mbox{\sc C\_Set\_Making}(u, \CSs, Rule(t))$;
        \STATE $\cs(u)\coloneqq\vs(u)\cup \ins(u)$;
        \STATE $\text{height}_T(u)\coloneqq\max_{1\leq i\leq m} \left\{\text{height}_{T}(c_{i})\right\} + 1$;
        \STATE $V\coloneqq V\setminus\{u\}$;
      \ENDIF
    \ENDFOR
  \ENDWHILE\label{match:endinductive}
  \STATE Let $r$ be the root of $t$ and $R$ the root of $T$;
  \IF{$r \in \cs(R)$}
    \RETURN ``yes'';~~/* $t$ matches $T$ */
  \ELSE
    \RETURN ``no'';
  \ENDIF
  \end{algorithmic}
  }
\end{algorithm}

In \cite{shoudai-ieice2018}, we proposed a polynomial-time algorithm for solving the membership problem for unordered term tree patterns with ordinal variables (i.e., variables without height constraints). The algorithm employs dynamic programming and runs in $O(nN^{3/2})$ time, where $n$ and $N$ denote the number of vertices in a given unordered term tree pattern $t$ and an unordered tree $T$, respectively.
When height-constrained (HC) variables are introduced, additional computation is required to check whether each substitution satisfies the corresponding height constraints. In this section, we present a polynomial-time algorithm for the Membership Problem for $\luttp_{\Sigma,\Lambda, X^{\h}}$ under the assumption that $|\Sigma| = |\Lambda| = 1$. This algorithm can be easily adapted to handle larger alphabets $\Sigma$ and $\Lambda$, making it applicable to more general settings.

Let $T$ be an unordered tree in $\ut_{\Sigma,\Lambda}$ and $t$ an unordered term tree pattern in $\luttp_{\Sigma,\Lambda,X^{\h}}$.  
We say that $t$ \emph{matches} $T$ if $T \in L(t)$.  
For a vertex $u'$ in $t$, we denote by $t[u']$ the unordered term tree pattern induced by $u'$ and all its descendants.  
The same notation applies to unordered trees; that is, for a vertex $u$ in $T$, $T[u]$ denotes the subtree rooted at $u$.
For vertices $u$ and $v$ in $T$ with $v$ a descendant of $u$, we define $T[u] - T[v]$ as the subtree rooted at $u$, excluding all proper descendants of $v$. Note that the vertex $v$ itself remains in $T[u] - T[v]$.
Let $\text{height}(T)$ denote the height of $T$, and for a vertex $u$ in $T$, let $\text{height}_T(u)$ be the height of $T[u]$, and $\overline{\text{height}}_T(u)$ be the height of $T - T[u]$.  
For two vertices $u$ and $v$ in an unordered tree or an unordered term tree pattern, we denote by $\text{dist}(u, v)$ the length of the path between them.

Let $u'$ be the child port of an $(i,j)$-HC-variable in $t$.  
We present Algorithm~\ref{proc:Matching} to solve the Membership Problem for $\luttp_{\Sigma,\Lambda,X^{\h}}$.  
The algorithm assigns to each vertex $c$ in $T$ a triple $(u', i', j')$, where $0 \leq i' \leq i-1$ and $0 \leq j' \leq j$.  
Such a triple indicates that there exists a proper descendant $v$ of $c$ such that $t[u']$ matches $T[v]$, and $j' = \overline{\text{height}}_{T[c]}(v)$.
In general, $i'$ may exceed $i$, and multiple values of $j'$ may correspond to the same $u'$ and $i'$, depending on the choice of $v$.  
However, we observe the following:
(1) We do not need to store triples with $j' > j$, as they exceed the height constraint, (2) the value $i' = i - 1$ is sufficiently large for use by any proper ancestor of $c$; larger values are unnecessary, and (3) for fixed values of $u'$ and $i'$, it is sufficient to store the minimum value of $\overline{\text{height}}_{T[c]}(v)$ among all matching descendants $v$.
Based on these observations, we define two sets for each vertex $u \in V_T$ that store only the information necessary for the algorithm.

\smallskip

\begin{definition}[Correspondence set]
For a vertex $u\in V_{T}$, a set $B \subset (V_{t}\times\mathbb{N}\times\mathbb{N}) \cup V_{t}$ is said to be the {\it correspondence set\/} for $u$ with respect to $t$ if the following Condition (C) holds:

{\bf Condition~(C)\/} on $u$: $\left((u',0,0)\in B\text{ and }u'\text{ is a child port}\right)$ or $\left(u'\in B\text{ and }u'\text{ is not a child port}\right)$ if and only if $t[u']$ matches $T[u]$. That is, the correspondence set for $u$ with respect to $t$ is
\begin{align*}
\vs(u) = & \left\{\!\!\!
\begin{array}{l|l}
(u',0,0) & \begin{array}{l}
u'\text{ is a child port of }t,\\
t[u']\mbox{ matches }T[u]
\end{array}
\end{array}\!\!\!\right\}\\
& \cup
\left\{\!\!\!
\begin{array}{l|l}
u' & \begin{array}{l}
u'\text{ is not a child port of }t,\\
t[u']\mbox{ matches }T[u]
\end{array}
\end{array}\!\!\!\right\}.
\end{align*}
\end{definition}

\smallskip

\begin{definition}[Inheritance set]
For a vertex $u\in V_{T}$, a set $B$ of triples $(u',i',j')\in V_{t}\times\mathbb{N}\times\mathbb{N}$ is said to be the {\it inheritance set\/} for $u$ with respect to $t$ if the following Condition (I) holds:

{\bf Condition~(I)\/} on $u$: For the child port $u'$ of any $(i,j)$-HC-variable of $t$ for some integers $1\leq i\leq j$, the triple $(u',i',j') \in B$ if and only if one of the following holds:
\begin{enumerate}
\item If $i' < i - 1$, then $I_1(u,u',i')\neq\emptyset$ and
$$
j' = \min\left\{\,\overline{\text{height}}_{T[u]}(v) \mid v\in I_1(u,u',i')\,\right\},
$$
where
$$
I_{1}(u,u',i') = 
\!\left\{\!\begin{array}{l|l}
  \!v\in V_{T} &
  \!\begin{array}{l}
  v \text{ is a descendant of } u,\\
  \text{dist}_{T}(u,v)=i',\\
  \overline{\text{height}}_{T[u]}(v) \leq j,\text{ and}\\
  (u',0,0)\in \vs(v)
\end{array}\!
\end{array}\!
\right\}\!.
$$
\item If $i' = i - 1$, then $I_2(u,u')\neq\emptyset$ and
$$
j' = \min\left\{\,\overline{\text{height}}_{T[u]}(v) \mid v\in I_2(u,u')\,\right\},
$$
where
$$
I_2(u,u')=
\!\left\{\!\begin{array}{l|l}
\!v\in V_{T} &
\!\begin{array}{l}
v \text{ is a descendant of }u,\\
\text{dist}_{T}(u,v)\geq i-1,\\
\overline{\text{height}}_{T[u]}(v) \leq j,\text{ and}\\
(u',0,0)\in \vs(v)
\end{array}\!
\end{array}\!
\right\}\!.
$$
\end{enumerate}
The inheritance set for $u$ with respect to $t$ is denoted by $\ins(u)$.
\end{definition}

\smallskip

Let $\cs(u)$ denote the union of the correspondence set $\vs(u)$ and the inheritance set $\ins(u)$:
$$
\cs(u) = \vs(u) \cup \ins(u).
$$

\subsection{Procedure {\sc I\_Set\_Making}}\label{subsec:i-rule}

Procedure {\sc I\_Set\_Making}, utilized as an auxiliary procedure in Algorithm~\ref{proc:Matching}, is identical to the one for ordered term tree patterns with height constraints presented in~\cite{shoudai-ieice2017}.
Therefore, we omit its details here and refer the reader to~\cite{shoudai-ieice2017} for a complete description.

An outline of Procedure {\sc I\_Set\_Making\/} is as follows:
Let $u$ be an internal vertex of $T$, and let $c_1,\ldots,c_m$ be all children of $u$.
Let $u'$ be the child port of an $(i,j)$-HC-variable of $t$ for some integers $1\leq i\leq j$.
For each integer $i'$ with $0\leq i'\leq i-1$, let $c_{m_1},\ldots,c_{m_k}$ be all children of $u$ such that $(u',i',j'_\ell)\in\cs(c_{m_\ell})$ for some integer $j'_\ell$ $(0\leq j'_\ell\leq j, 1\leq\ell\leq k)$.
Define $M_1$ and $M_2$ as indices $(1\leq M_1,M_2 \leq m)$ satisfying the following conditions:
\begin{itemize}
\item The subtree $T[c_{M_1}]$ has the maximum height among the subtrees $\{T[c_1],\ldots,T[c_m]\}$
\item The subtree $T[c_{M_2}]$ has the maximum height among $\{T[c_1],\ldots,T[c_m]\}\setminus\{T[c_{M_1}]\}$.
\end{itemize}
Then, there is a descendant $v$ of $u$ satisfying $(u',0,0)\in\cs(v)$, $\text{dist}(u,v)=i'+1$, and $\overline{\text{height}}_{T[u]}(v)=j(i')$, where
$$
j(i')=
\left\{\!\!
\begin{array}{ll}
\multicolumn{2}{l}{
\min\left\{\!\!
\begin{array}{l}  
\text{height}_{T[u]}(c_{M_1}),\\
\max\{j_{M_1},\text{height}_{T[u]}(c_{M_2})\}
\end{array}\!\!
\right\}+1
}\\
& \text{if }M_1\in\{m_1,\ldots,m_k\},\\
\text{height}_{T[u]}(c_{M_1})+1\hspace{22pt} & \text{otherwise.}
\end{array}
\right.
$$
For each $i'$ $(0\leq i' < i-2)$, if $j(i')\leq j$, we add $(u',i'+1,j(i'))$ to $\ins(u)$.
Moreover, if $\min\{j(i-2),j(i-1)\}\leq j$, we add $(u',i-1,\min\{j(i-2),j(i-1)\})$ to $\ins(u)$.
Procedure {\sc I\_Set\_Making\/} for an internal vertex $u$ of $T$ is defined by performing these computations for the vertex $u$
over all child ports $u'$ of $t$.
It can be shown that the set $\ins(u)$ obtained by this procedure correctly satisfies Condition~(I).
Therefore, $\ins(u)$ is the inheritance set for $u$ with respect to $t$.


 Algorithm {\sc Match-$\luttp_{\Sigma,\Lambda,X^{\h}}$} (Algorithm \ref{proc:Matching}) computes the inheritance sets $\ins(u)$ and the correspondence sets $\vs(u)$ for all vertices $u$ of a given unordered tree $T$, proceeding from the leaves to the root.
In the next subsection, we introduce a new algorithm, Procedure {\sc C\_Set\_Making} (Procedure~\ref{proc:c-set-making}), for computing the correspondence sets with respect to an unordered term tree pattern.

\subsection{Procedure {\sc C\_Set\_Making}}\label{subsec:c-rule}
\algsetup{linenosize=\small}
\begin{algorithm}[t]
  \floatname{algorithm}{Procedure}
  \caption{\sc C\_Set\_Making}\label{proc:c-set-making}
  {\fontsize{9}{10.5}\selectfont 
  \begin{algorithmic}[1]
  \REQUIRE a vertex $u$ of $T$, the set of all $\CSs$ for all children of $u$, and a set $Rule$ of matching rules;
  \ENSURE a correspondence set $\vs$ of $u$;
  \smallskip
  \STATE $\vs :=\emptyset$;
  \STATE Let $c_{1},\ldots,c_{m}$ be all children of $u$;
  \FOR{each $rule = (u'\leftarrow J(c'_{1}),\ldots,J(c'_{m'}))$ in $Rule$} \label{CSetMaking_For_Start}
    \STATE $V \coloneqq \{c_{1},\ldots,c_{m}\}$,~$V'\coloneqq \{c'_{1},\ldots,c'_{m'}\}$;
    \STATE $E \coloneqq \emptyset$;
    \STATE $P\coloneq \{c' \mid c' \in V'\text{ and }c'\text{ is a child port}\}$;
    \STATE $h_{\max} \coloneqq \max\left(\{0\}\cup\{j \mid J(c') = (c',i,j) \text{ for } c' \in P\}\right)$;
    \FOR{each $c'_{\ell'}$ $(1\leq \ell' \leq m')$}
      \FOR{each $c_{\ell}$ $(1\leq \ell \leq m)$}
        \IF{$c'_{\ell'}$ is a child port}
          \STATE $i \coloneqq\text{ the second component of }J(c'_{\ell'})=(c'_{\ell'},i,j)$;
          \IF{$(c'_{\ell'},i-1,j)\in \cs(c_{\ell})$ for some $j$}
            \STATE $E\coloneqq E\cup\{(c_{\ell},c'_{\ell'})\}$;
          \ENDIF
        \ENDIF
        \IF{$c'_{\ell'}$ is not a child port and $c'_{\ell'}\in\cs(c_{\ell})$}
          \STATE $E\coloneqq E\cup\{(c_{\ell},c'_{\ell'})\}$;
        \ENDIF
      \ENDFOR
    \ENDFOR
    \IF{$m > m'$}
      \STATE Let $d_{1},\ldots,d_{m-m'}$ be new $m-m'$ vertices;
      \STATE $V' \coloneqq V'\cup \{d_{1},\ldots,d_{m-m'}\}$;
      \FOR{$d \in \{d_{1},\ldots,d_{m-m'}\}$}
        \FOR{each $c_{\ell}$ $(1\leq \ell \leq m)$}
          \IF{$\text{height}_{T[u]}(c_{\ell}) < h_{\max}$}
            \STATE $E\coloneqq E\cup\{(c_{\ell},d)\}$;
          \ENDIF
        \ENDFOR
      \ENDFOR
    \ENDIF
    \STATE $G \coloneqq (V\cup V',E)$;
    \IF{$G$ has a perfect graph matching}
      \IF{$u'$ is a child port}
        \STATE $\vs \coloneqq \vs \cup \{(u',0,0)\}$;
      \ELSE
        \STATE $\vs \coloneqq \vs \cup \{u'\}$;
      \ENDIF
    \ENDIF
  \ENDFOR\label{CSetMaking_For_End}
  \RETURN $\vs$;
  \end{algorithmic}
  }
\end{algorithm}

We define Procedure {\sc C\_Set\_Making} (Procedure~\ref{proc:c-set-making}), which computes the correspondence set $\cs(u)$ for an internal vertex $u$ of an unordered tree $T$, based on the correspondence sets of its children and a set of matching rules defined over an unordered term tree pattern $t$.

\smallskip

\begin{definition}[Matching rules]
Let $u'$ be an internal vertex in $t$, and let $c'_1,\ldots,c'_{m'}$ be its children.
The {\it matching rule\/} for $u'$ is defined as:
$$
u'\leftarrow J(c'_{1}),\ldots,J(c'_{m'}),
$$
where for $\ell'=1,\ldots,m'$,
$$
J(c'_{\ell'})=\left\{\begin{array}{ll}
c'_{\ell'} & \begin{array}{l}
\text{if $c'_{\ell'}$ is not a child port,}
\end{array}\\
(c'_{\ell'},i,j) & \begin{array}{l}
\text{if $c'_{\ell'}$ is a child port of an $(i,j)$-}\\
\text{HC-variable.}
\end{array}
\end{array}\right.
$$
The set of all such rules for internal vertices in $t$ is denoted by $Rule(t)$.
\end{definition}

\smallskip

The procedure determines whether a rule $u' \leftarrow J(c'_1), \ldots, J(c'_{m'})$ is applicable to $u$ by constructing a bipartite graph $G = (V \cup V', E)$, where $V = \{c_1, \ldots, c_m\}$ are the children of $u$, and $V' = \{c'_1, \ldots, c'_{m'}\}$ are the children of $u'$.  
An edge $(c_\ell, c'_{\ell'})$ is added to $E$ if
$c'_{\ell'}$ is a child port and $(c'_{\ell'}, i-1, j) \in \cs(c_\ell)$ for some $j$, or $c'_{\ell'}$ is not a child port and $c'_{\ell'} \in \cs(c_\ell)$.

If $m > m'$, $m - m'$ dummy vertices $d_1, \ldots, d_{m - m'}$ are added to $V'$ to balance the two sides. Let $h_{\max}$ be the maximum height constraint among the child ports in $\{c'_1, \ldots, c'_{m'}\}$. Each dummy vertex is connected to any $c_\ell$ such that the height of $T[c_\ell]$ is less than $h_{\max}$.

This construction ensures that:
\begin{itemize}
\item Any subtree $T[c_\ell]$ with height $\geq h_{\max}$ cannot be fully absorbed by an HC-variable and must be matched to a corresponding vertex in $t$.
\item Any subtree with height $< h_{\max}$ may be absorbed into an HC-variable and matched via a dummy vertex.
\end{itemize}

If the bipartite graph $G$ has a perfect matching, the rule is valid, and $u'$ is added to $\cs(u)$ as $(u', 0, 0)$ if it is a child port, or as $u'$ otherwise.

\begin{figure}[t]
  \begin{center}
  \includegraphics[scale=0.35]{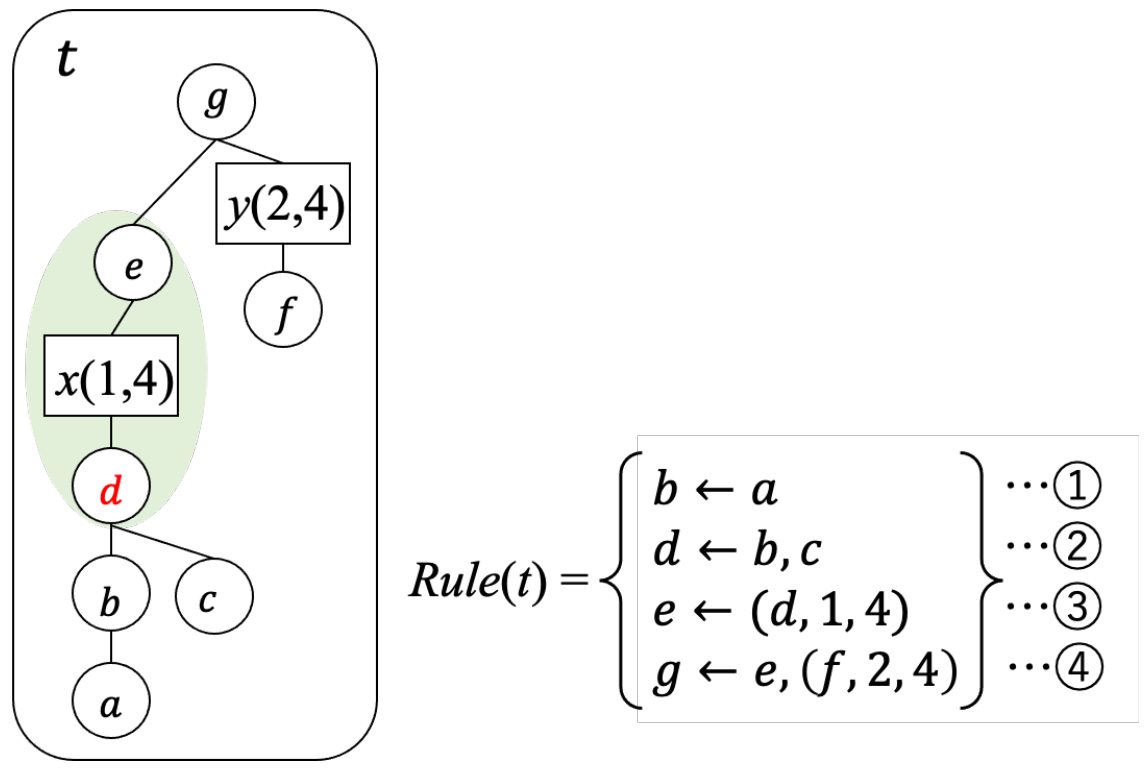}\\
  \includegraphics[scale=0.4]{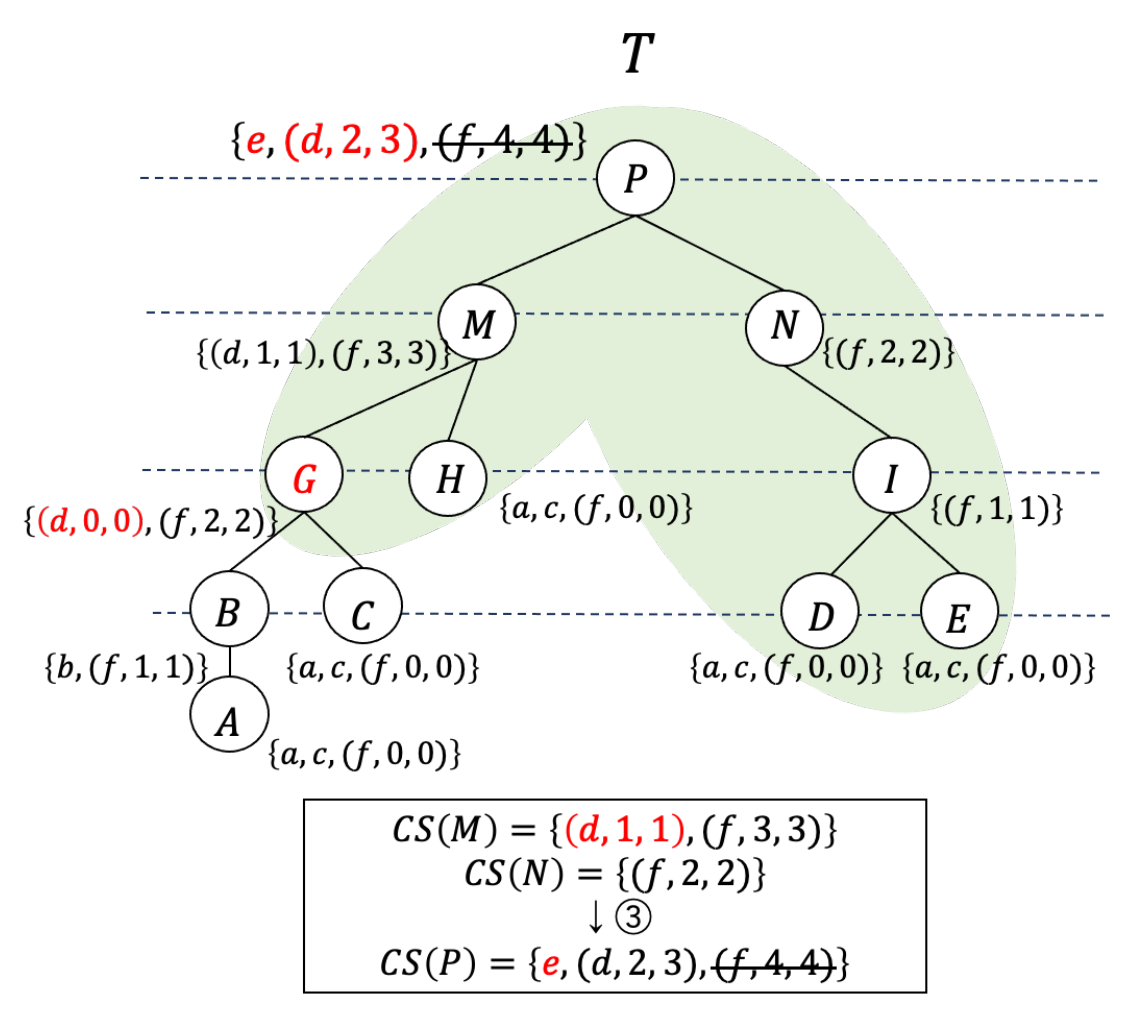}
  \end{center}
  \caption{
    Construction of matching rules $Rule(t)$ from an unordered term tree pattern $t$ and their application to an unordered tree $T$:
    In this example, the root $g$ of $t$ does not appear in $\cs(P)$ of $T$.
    Therefore, $t$ does not match $T$.
  }\label{fig:cset-example}
\end{figure}

Fig.~\ref{fig:cset-example} shows an example execution of Procedure {\sc C\_Set\_Making\/} applied to all vertices of an unordered tree $T$.
In this example, a matching rule from $Rule(t)$ is used to compute $\cs(P)$ for a vertex $P$ in $T$.

\subsection{Time Complexity Analysis}

We first analyze the maximum number of elements in $\cs(u)$ for $u \in V_T$.  
Each $\cs(u)$ depends on the number of vertices $n$ in $t$ and the trunk constraints $i(u')$ for each child port $u' \in V_t$.  
The total number of elements is at most
$
n + \sum_{\text{child port } u' \in V_{t}} i(u').
$
Let $\mathcal{S} = \sum_{\text{child port } u' \in V_t} i(u')$.  
Then, the size of $\cs(u)$ is bounded by $O(\max\{n, \mathcal{S}\})$.

The height of each vertex in $T$ can be computed in $O(N)$ time.  
To compute, for each vertex, the maximum height among its siblings (excluding itself), we apply a sorting-based method.  
For each parent, we sort its children and compute sibling maxima, resulting in a total cost of
$
\sum_{u \in V_{T}} O(d_{u} \log d_{u}) = O\left(\log D \sum_{u \in V_{T}} d_{u}\right) = O(N \log D),
$
where $d_u$ is the number of children of $u$, and $D$ is the maximum degree in $T$.

Next, we analyze the cost of applying matching rules.  
These rules are applied to each internal vertex $u'$ of $t$, requiring $O(n)$ total applications.  
For each $u'$, we construct a bipartite graph between the children of $u'$ and those of a vertex $u$ in $T$, adding dummy vertices if necessary.  
Let $d_{u'}$ be the number of children of $u'$.
Therefore, each bipartite graph needs
$O(d_{u'} \cdot \max\{n, \mathcal{S}\}) + O((d_{u} - d_{u'}) \cdot \max\{n, \mathcal{S}\}) = O(d_{u} \cdot \max\{n, \mathcal{S}\})$ time for its construction.
Applying this over all vertices $u$ in $T$ yields a total cost of $O(N \cdot \max\{n, \mathcal{S}\})$.

For inheritance rules, the algorithm checks whether $(c, i, j)$ can be inherited at each vertex $v$ in $T$.  
If the trunk condition is not satisfied, the new trunk value becomes $i+1$, and the height becomes $\max\{\text{height}_T(u), i\} + 1$.  
Inheritance is not performed if the trunk exceeds $i-1$.  
These checks take constant time per vertex, yielding $O(\max\{n, \mathcal{S}\})$ per vertex and $O(N \cdot \max\{n, \mathcal{S}\})$ in total.

Let $T_{\rm GM}(d_u)$ denote the time for testing the existence of a perfect matching in a bipartite graph with degree $d_u$.  
Then the total cost for all matching rule applications is $O(n \cdot T_{\rm GM}(d_u))$.  
Using the Hopcroft--Karp algorithm, which runs in $O(d_u^{5/2})$ time, the total becomes $\sum_{u \in V_T} O(n \cdot d_u^{5/2}) \leq \sum_{u \in V_T} O(n \cdot d_u \cdot D^{3/2}) = O(n N D^{3/2})$.

Combining all components, the overall time complexity is $O(N \log D) + O(N \cdot \max\{n, \mathcal{S}\}) + O(n N D^{3/2}) = O\left(N \cdot \max\{n D^{3/2}, \mathcal{S}\}\right)$.
Thus, we obtain the following result. The formal proof of correctness is omitted due to space limitations.

\smallskip

\begin{theorem}\label{thm:member-luttp}
  For any sets of vertex labels $\Sigma$ and edge labels $\Lambda$, {\bf Membership Problem for $\luttp_{\Sigma,\Lambda, X^{\h}}$\/} is solvable in $O(N \cdot \max\{n D^{3/2}, \mathcal{S}\})$ time, where $n$ and $N$ are the numbers of vertices of a given unordered term tree pattern $t$ and a given unordered tree $T$, respectively, $\mathcal{S}$ is the sum of trunk constraints for all child ports in $t$, and $D$ the maximum number of the children of an internal vertex of $T$.
\end{theorem}

\subsection{Computational Experiments}\label{subsec:experiments}

\begin{figure}[t]
  \centering
  \includegraphics[scale=0.42]{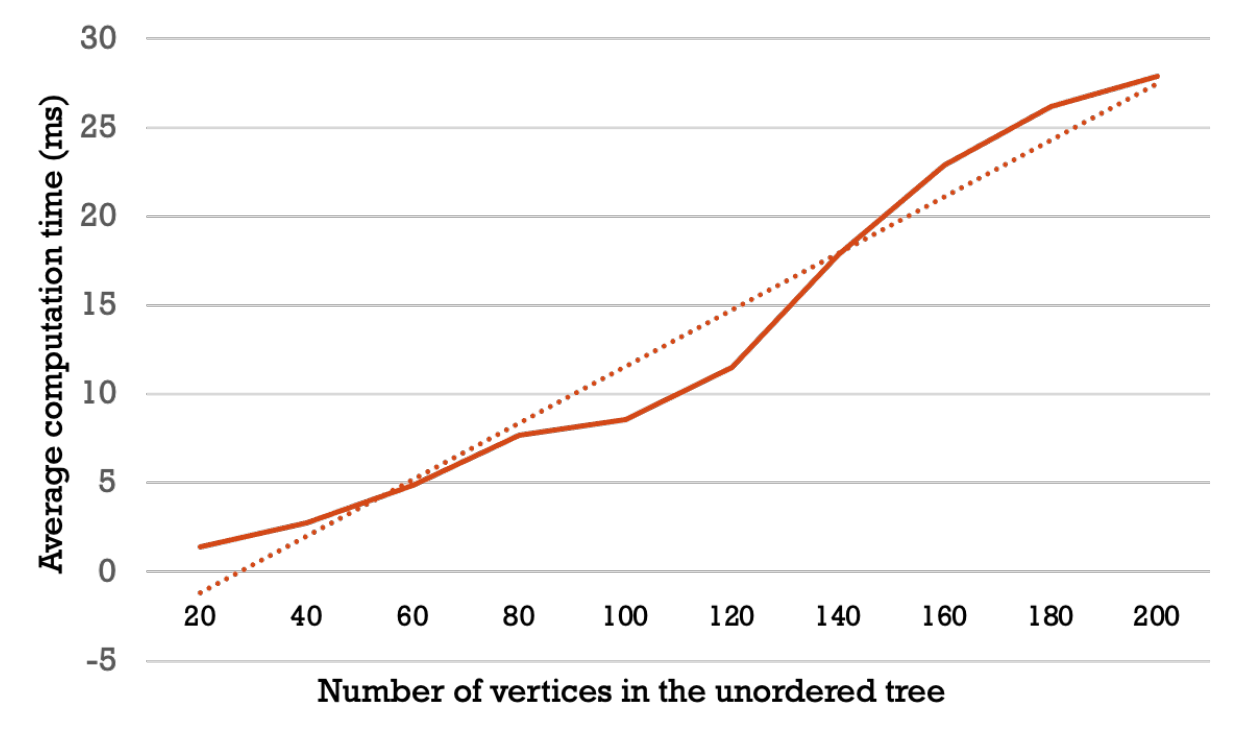}
  \vspace{1em} 
  \begin{scriptsize}
  \begin{tabular}{|c||ccccc|}
  \hline
  \textbf{Number of Vertices} & 20 & 40 & 60 & 80 & 100 \\
  \textbf{Avg. Time (ms)} & 1.395 & 2.757 & 4.849 & 7.687 & 8.562 \\
  \hline
  \multicolumn{6}{c}{} \\[-0.8em]
  \hline
  \textbf{Number of Vertices} & 120 & 140 & 160 & 180 & 200 \\
  \textbf{Avg. Time (ms)} & 11.478 & 17.821 & 22.909 & 26.163 & 27.896 \\
  \hline
  \end{tabular}
\end{scriptsize}
\caption{Computation time increases approximately proportionally to the number of vertices in the unordered tree.}\label{fig:matching-time}
\end{figure}

We conducted computational experiments to evaluate the performance of our algorithm for matching unordered term tree patterns with height-constrained variables. In particular, we examined how the number of vertices in an unordered tree affects the overall matching time.

For this purpose, we prepared ten unordered trees that match a fixed unordered term tree pattern. The number of vertices in these trees ranged from 20 to 200, increasing in increments of 20. For each tree, the matching algorithm was executed 5,000 times, and the average computation time was recorded. These average values were then plotted to observe the overall trend.
All experiments were conducted on a machine equipped with an Intel(R) Core(TM) i7-10700 CPU @ 2.90GHz, 16 GB of RAM, and an NVIDIA GeForce RTX 3050 GPU. The algorithm was implemented in Python.

Fig.~\ref{fig:matching-time} shows the number of vertices and the corresponding average computation time (in milliseconds). As seen in the figure, the computation time increases with the number of vertices. Moreover, the relatively low average times suggest that matching unordered term tree patterns with height constraints can be performed efficiently, even for large trees.

\section{Concluding Remarks}\label{sec:concl}

To represent unordered tree-structured patterns with rich structural features, we propose unordered term tree patterns with height-constrained variables and develop a polynomial-time algorithm for solving the membership problem.

As a next step, we plan to apply the proposed algorithm to real unordered tree data, such as glycan structures, in order to evaluate its practical effectiveness in extracting meaningful patterns.

From the viewpoint of computational learning theory, we also study the polynomial-time learnability of unordered tree-structured patterns with isomorphism-invariant constraints, with particular focus on those involving height constraints.

\end{document}